\documentclass[11pt, titlepage]{article}
\pdfoutput=1

\usepackage{amssymb}
\usepackage{amsmath}
\usepackage{graphicx}
\usepackage{cleveref}
\usepackage{authblk}
\usepackage{caption}
\usepackage{subcaption}
\usepackage[title]{appendix}
\crefname{equation}{Eq.}{Eqs.}
\crefname{figure}{Fig.}{Figs.}
\crefname{tabular}{Tab.}{Tabs.}
\crefname{table}{Table}{Tables}
\newcommand{\un}[1]{\>\mathrm{#1}}

\usepackage{sectsty}
\sectionfont{\fontsize{12}{15}\selectfont}

\title{Laser-induced electron dynamics and surface modification in ruthenium thin films}
\date{}

\author[1,*]{Fedor Akhmetov}
\author[1]{Igor Milov}
\author[2]{Sergey Semin}
\author[2]{Fabio Formisano}
\author[3,4]{Nikita Medvedev}
\author[1]{Jacobus M. Sturm}
\author[5]{Vasily V. Zhakhovsky}
\author[1]{Igor A. Makhotkin}
\author[2]{Alexey Kimel}
\author[1]{Marcelo Ackermann}

\affil[1]{Industrial Focus Group XUV Optics, 
        MESA+ Institute for Nanotechnology,
        University of Twente, Drienerlolaan 5,
        Enschede, 7522 NB, The Netherlands}

\affil[2]{Institute for Molecules and Materials,
        Radboud University Nijmegen, Heyendaalseweg 135, 
        Nijmegen, 6525 AJ, The Netherlands}
            
\affil[3]{Institute of Physics,
        Czech Academy of Sciences,
        Na Slovance 1999/2,
        Prague 8, 182 21, Czech Republic}
            
\affil[4]{Institute of Plasma Physics,
        Czech Academy of Sciences, 
        Za Slovankou 3,
        Prague 8, 182 00, Czech Republic}

\affil[5]{Joint Institute for High Temperatures, RAS,
        Izhorskaya St. 13/2, Moscow, 125412, Russia}
        
\affil[*]{Corresponding author: f.akhmetov@utwente.nl}

\begin{document}

\maketitle

\begin{abstract}
We performed the experimental and theoretical study of the heating and damaging of ruthenium thin films induced by femtosecond laser irradiation. Results of an optical pump-probe thermoreflectance experiment with rotating sample allowing to significantly reduce heat accumulation in irradiated spot are presented. We show the evolution of surface morphology from growth of a heat-induced oxide layer at low and intermediate laser fluences to cracking and grooving at high fluences. Theoretical analysis of pump-probe signal allows us to relate behavior of hot electrons in ruthenium to the Fermi smearing mechanism. The analysis of heating is performed with the two-temperature modeling and molecular dynamics simulation, results of which demonstrate that the calculated melting threshold is higher than experimental damage threshold. We attribute it to heat-induced surface stresses leading to cracking which accumulates to more severe damage morphology. Our results provide an upper limit for operational conditions for ruthenium optics and also direct to further studies of the Fermi smearing mechanism in other transition metals.
\end{abstract}


\section{Introduction}
\label{sec:i}
Ultrashort lasers have become versatile tools for manipulation of material properties at the nanoscale. A unique ability of such lasers to deliver enormous amount of energy into a sample on a femtosecond timescale creates extreme strongly non-equilibrium states which upon relaxation lead to altered material properties. Ultrashort lasers are used for surface nanostructuring \cite{Bonse2012FemtosecondStructures} and nanofabrication \cite{Murakami2009Burst-modeAblation, Han2017ControllableIrradiation, Zhang2017LaserApplications}, as well as for reversible switching of material structure between crystal and amorphous phases for applications in data storage \cite{Yang2015ControllablePulses}. Alternatively to these high intensity laser applications, ultrashort lasers with relatively low intensity are used in various metrology and probing techniques \cite{Edward2020Laser-inducedRoughness, Zhao2021NanometerRetrieval, Oh2005FemtosecondMetrology}. In this context, contrarily to the previous examples, laser-induced damage must be avoided. In both scenarios of desired and undesired material modifications, a precise control of the output of laser-matter interaction is crucial, and can only be achieved with a deep understanding of the fundamental physical processes involved. This work aims at such understanding on an example of ruthenium (Ru) thin films exposed to optical femtosecond laser pulses.

Ru and Ru oxides are indispensable materials for various catalysis applications \cite{Over2012SurfaceResearch, Goodman2007COSurface}. Controlling the surface properties of Ru such as oxidation state, as well as the shape and structure of Ru nanocrystals, provides additional efficient functionalities \cite{Axet2020CatalysisNanoparticles}. Ultrathin films of Ru are used as protective capping layers in extreme ultraviolet optics, due to high transmissivity in EUV range and low surface oxidation \cite{Bajt2008PropertiesIrradiation}. Ru is also considered as a high Z material for grazing incidence hard X-ray optics \cite{Aquila2015FluenceMirrors}.

In our previous studies we focused on severe damage of Ru thin films in single- \cite{Milov2018, Milov2020SimilarityAnd, Milov2020Two-levelIrradiation} and multi-shot regimes \cite{Makhotkin2018DamageThreshold}, as well as on a long-term exposure of Ru at fluences significantly below the single-shot ablation threshold \cite{Makhotkin2018}. Such sub-threshold investigation of material degradation is challenging  since the processes involved are elusive to be detected post-mortem. Dynamical data must be collected to provide insights into how laser-induced evolution of Ru results in final damage. Therefore, in this work we continue investigating Ru interaction with ultrashort laser pulses in an all-optical pump-probe scheme with rotating sample to reduce heat accumulation effects. After the irradiation, Ru surfaces are examined with various surface-sensitive characterization techniques.

To interpret experimental results we perform theoretical analysis of Ru optical response to ultrashort laser irradiation. When an ultrafast laser pulse illuminates a metallic target, its energy is absorbed by the conduction band electrons, which leads to evolution of initially equilibrium electron distribution to a non-equilibrium one. It is often assumed that the thermalization of an electron gas to an equilibrium Fermi distribution occurs on a timescale of $\sim 100$ fs, which is comparable to a pulse duration \cite{Fann1992ElectronGold, Sun1994Femtosecond-tunableGold, Wellershoff1999, Rethfeld2004TimescalesExcitation}. Thus, it is convenient to consider the probed optical response of a metallic target in terms of the electron temperature $T_e$ elevated with respect to the atomic one $T_a$. However, a reliable model for $T_e$-dependent optical constants is required for a direct analysis of optical pump-probe experiments. The widely used Drude model is limited in applicability to the case of simple metals \cite{Rethfeld2017ModellingAblation}. In the case of noble metals, the optical response can be successfully described by Fermi smearing mechanism, assuming that reflectivity change is proportional to the derivative of the Fermi distribution with respect to the electron temperature \cite{Lynch1972ThermomodulationCu, Schoenlein1987FemtosecondMetals, DeHaan2020}. 

In contrast, in a transition metal with a complex band structure such as Ru, the optical constants are formed by a sum of \textit{inter}- and \textit{intraband} contributions within a combined d-s/p conduction band and to date could not be reasonably approximated with a simple analytical model. Instead, $T_e$-dependence may be extracted from first-principles simulations of the complex dielectric function of a material in, e.g., random-phase approximation (RPA) \cite{Gunnarsson1989AbMetals, Gurtubay2005Electron-holeMeasurements, Ambrosch-Draxl2006,  Cazzaniga2010AbMetals}.

We test to what extent a simple Fermi smearing analysis can be applied to Ru. Absorption of the laser energy by the electrons and coupling of electrons to the lattice is modelled with the two-temperature model (TTM). The calculated melting threshold is compared with the experimental surface modification observations.

\section{Experimental setup}
\label{sec:es}

\begin{figure}[h!]
    \centering
    \includegraphics[height=0.3\paperheight]{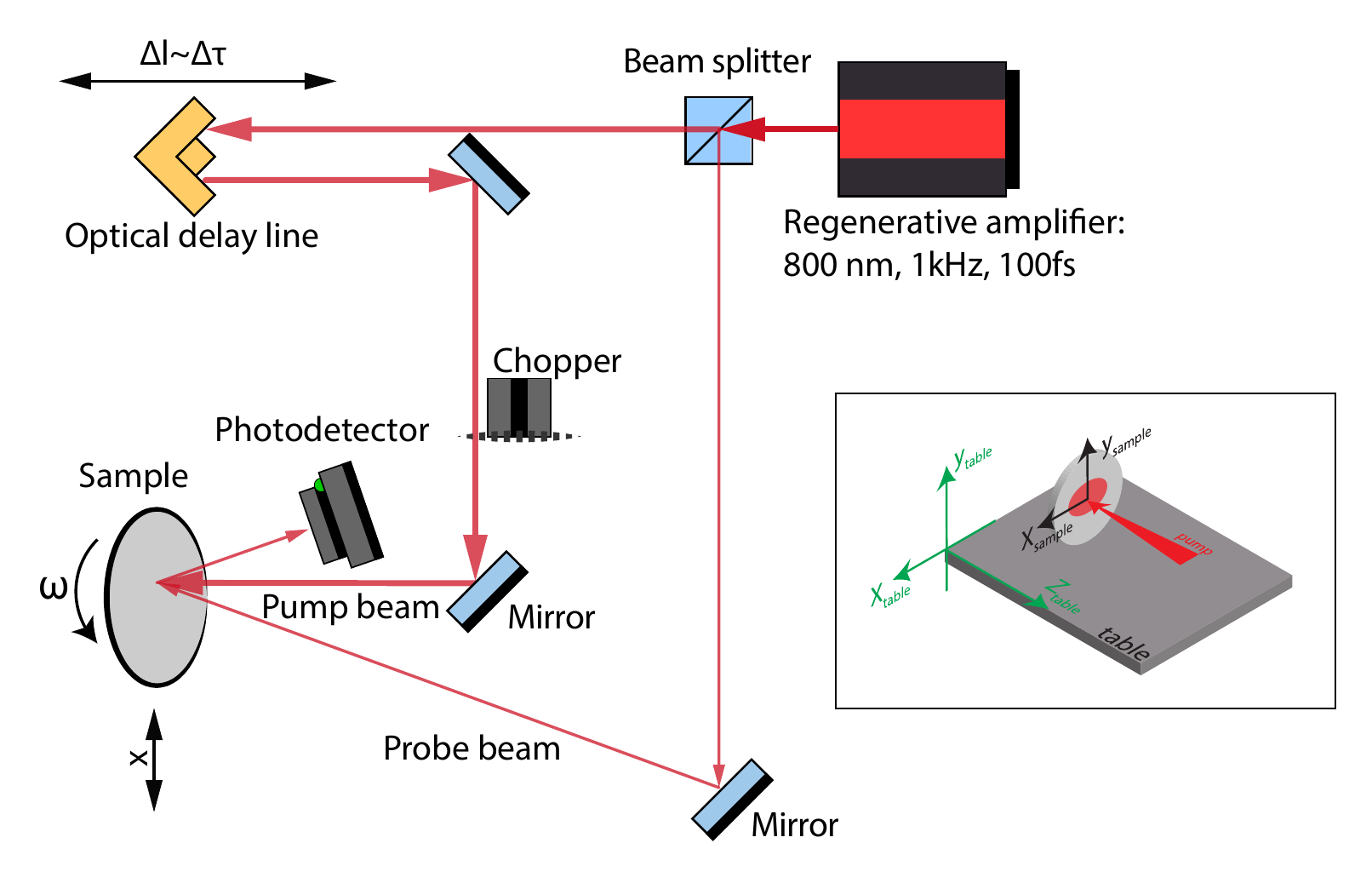}
    \caption{Scheme of experimental set-up used for optical pump-probe thermoreflectance measurements. The laser beam coming out of regenerative amplifier (Spitfire), which generates the near transform-limited 100 fs FWHM pulses with 800 nm wavelength ($\hbar\omega = 1.55\un{eV}$) and 1 kHz repetition rate. Laser beam is split into a high-intensity pump and a low-intensity probe pulses. The pump pulse is chopped to a 500 Hz repetition rate. The probe pulse at 1 kHz arrives to the sample with a time delay controlled with an optical delay line to measure the reflectivity change induced by the pump pulse. Inset: configuration of polarisation of incident beams and convention on axes.}
    \label{fig:exp}
\end{figure}


For the pump-probe thermoreflectance experiments, we employed pump-probe set-up based on ultrafast 1 kHz repetition rate Ti:Sapphire laser \cref{fig:exp}. The experiment was carried out under atmospheric conditions. The beam profile was characterized by a knife-edge method along the horizontal direction. Additional post mortem analysis of Ru ablation craters revealed an ellipticity of a Gaussian profile. The value of the semi-major axis was $\sim 115 \un{\mu m}$ (@ $1/e^2$) whereas the value of the semi-minor axis was 1.4 times smaller. The peak incindent fluence was calculated as follows:
\begin{equation}
\label{eq:1}
    F = \frac{2 E_{pulse}}{\pi w_x w_y}
\end{equation}
Here $E_{pulse}$ is an energy value of pump pulse, $w_x$ and $w_y$ are values of semi-major and semi-minor axes of ellipse, respectively. The sample was positioned slightly before the focal spot to avoid air ionization and consequent aberration of the beam quality at the sample. The angles of incidence (AOI) were set close to normal ($\sim 5\pm 2^{\circ}$ and $\sim 8\pm 2^{\circ}$ off-sample normal for pump and probe pulses, respectively). To continuously control the laser fluence we used an attenuator consisting of a half-wave plate and a polarizer located in the pump path. We characterized the laser pulse duration by placing an autocorrelator just before the sample with typical measured values to be $\sim 85 \un{fs}$ (FWHM). P- and s-polarizations with respect to the optical bench were used for the pump and the probe pulse, respectively.\par
As a studied material, we used ruthenium (Ru) polycrystalline thin metal film. The Ru films of various thickness between 17 and 125 nm were deposited by magnetron spattering on top of (100) single-crystal Si substrates wafers with 3-inch diameter. Ru thickness was measured using X-ray reflectivity technique. To reduce the effect of heat accumulation, samples were mounted on a rotational stage. The rotational frequency was set to $\omega=90 \un{Hz}$. Such a scheme ensured an effective reduction of the repetition rate, without changes of the laser source. However, after several rotations laser pulses start hitting previously exposed surface, therefore accumulation of irreversible changes induced by pump pulses is expected at high pump fluences. For each sample we measured transient reflectivity change induced by pump pulses of various fluences. For each fluence value a different position on a sample was measured. Since the position variation was small compared to the radius at which the set of measurements was performed, a slightly different level of accumulation of irreversible changes for different fluences within each sample can be neglected.\par
Before the actual measurement of thermoreflectance curve, several pump and probe pulses arrive to the sample surface at negative delays between the pump and the probe resulting in a weak signal. We have a pronounced step in this signal at negative delays from -15 to 0 ps (see \cref{fig:pumpprobe}). We explain it by a partial split of the pump signal in the beamsplitter: a small part of the pump comes to the sample just before the next probe and serves as a weak excitation. Analyzing thermoreflectance data, we aligned the signal to the values taken from delays $t<-15 \un{ps}$.

\section{Experimental results}
\label{sec:r}
\subsection{Fluence-dependent transient thermoreflectance}
\label{subsec:r1}
\begin{figure}[h]
    \centering
    \includegraphics[height=0.2\paperheight]{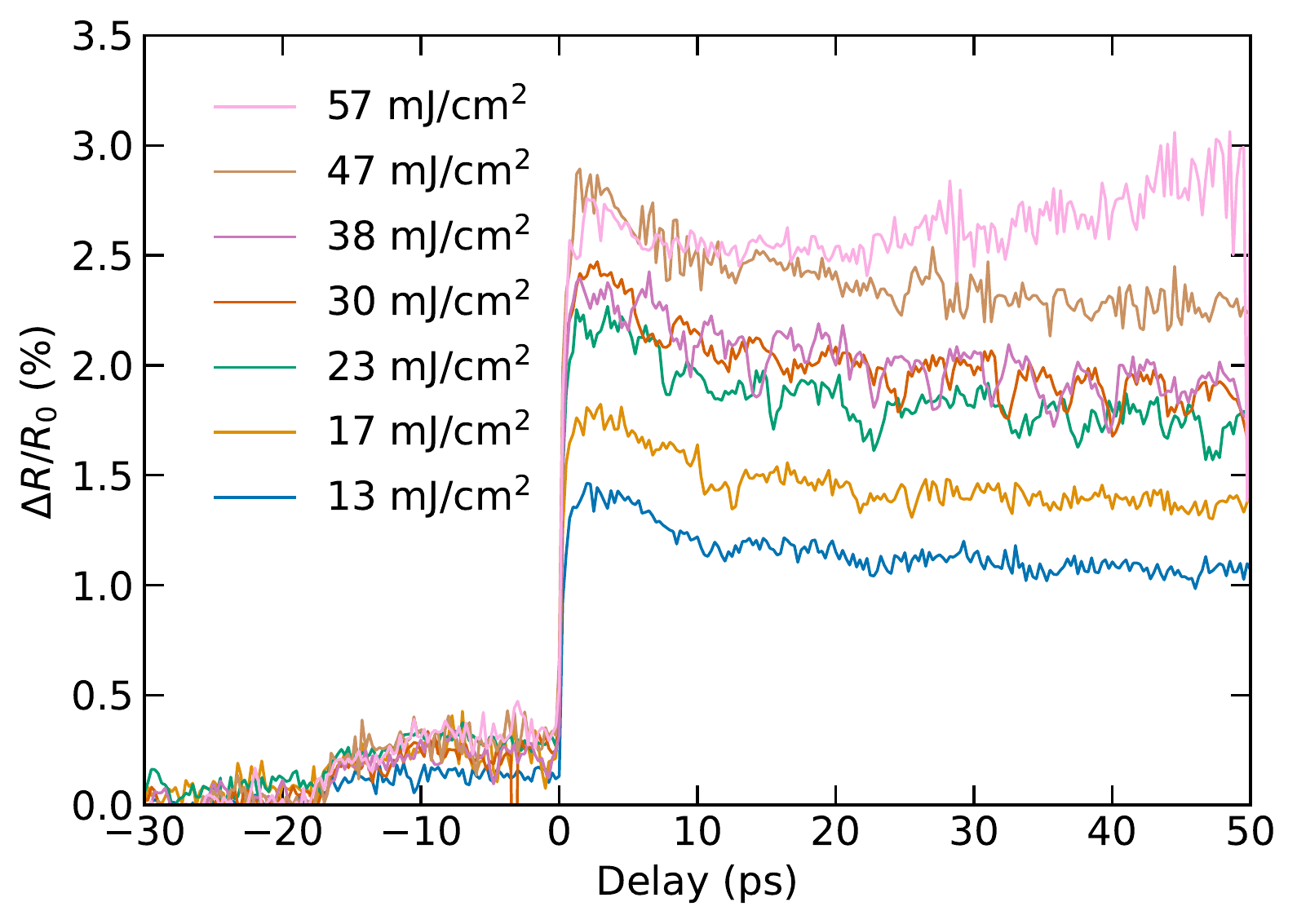}
    \caption{Transient changes in reflectivity of a 37 nm Ru film measured for various incident fluences.}
    \label{fig:pumpprobe}
\end{figure}

\begin{figure}[h]
    \centering
    \includegraphics[height=0.2\paperheight]{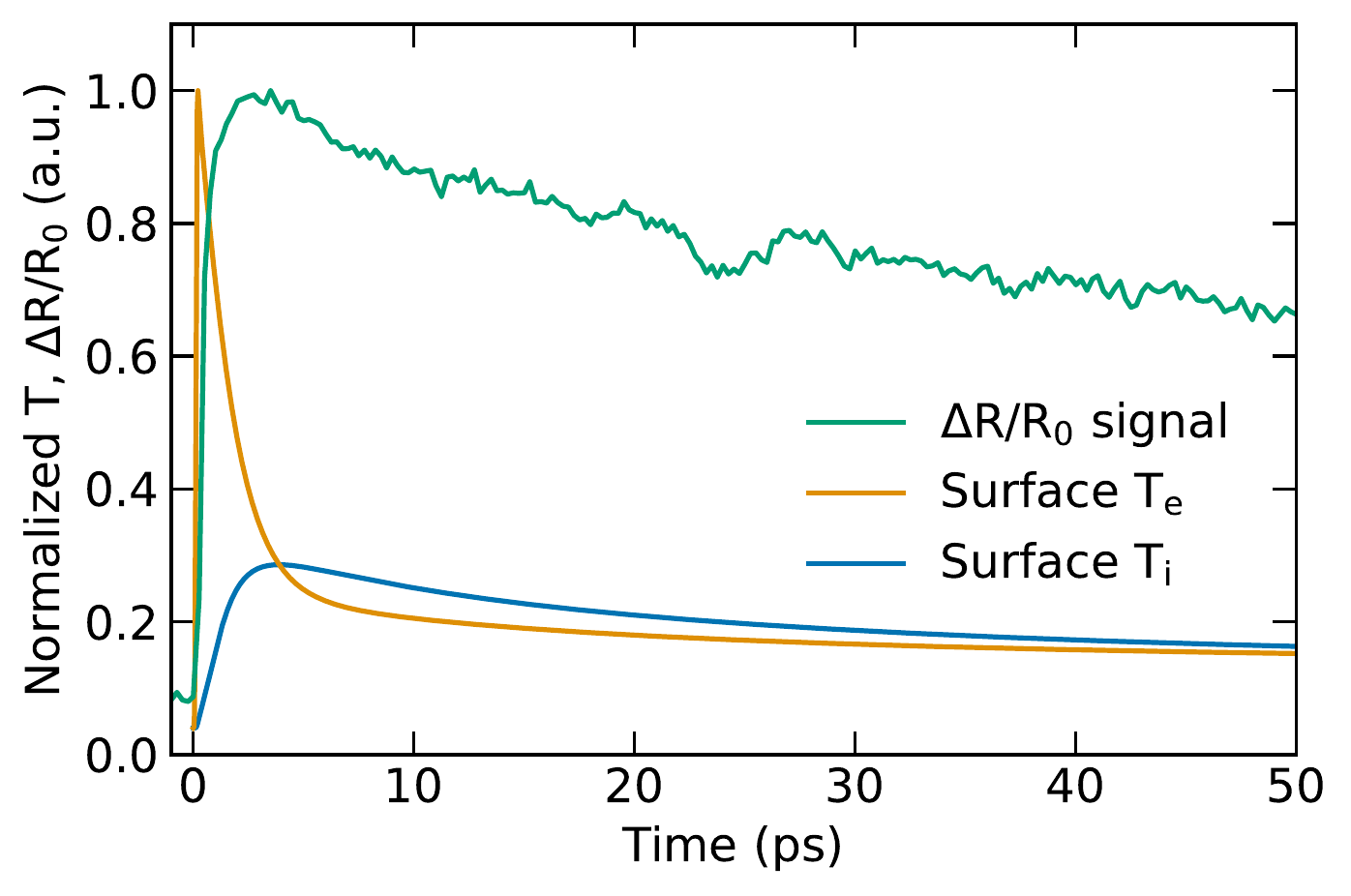}
    \caption{Normalized pump-probe reflectivity signal (green) compared with normalized profiles of electron (orange) and lattice (blue) temperatures.}
    \label{fig:prof}
\end{figure}

A set of pump-probe transient reflectivity data for 37 nm thick Ru film is shown in \cref{fig:pumpprobe}. It demonstrates that in all measured curves the reflected probe intensity sharply increases during the first few ps resulting in a pronounced peak followed by a slow decay. The very sharp increase at $<1 \un{ps}$ timescales is primarily associated with the increase of the electron temperature induced by the absorbed pump pulse. At $\sim 1-3 \un{ps}$ timescale, a slower signal increase can be attributed to the coupling of the electrons to the lattice and thereby induced lattice heating. A slow decay over tens of picoseconds corresponds to heat diffusion from the surface region deeper into the sample \cite{Hohlfeld2000}. This is confirmed by the TTM simulations of the surface electron and lattice temperatures shown in \cref{fig:prof}. \par 

At high pump pulse fluences above $57 \un{mJ/cm^2}$ we observe qualitatively different behavior of $\Delta R/R_0$ signal at $\Delta t>20 \un{ps}$ time delay compared to lower fluences. We see the change of trend from a slow decay to an increase. For every thickness, we were able to extract the time when signal starts to increase by calculating time derivative of a signal using regularization algorithm for noisy signals \cite{Chartrand2011NumericalData}. We determined the time and fluence values when a derivative of a signal changes its sign from negative to positive, and stored them in \cref{tab:1}. For 17 nm Ru film, we could not extract any values, because the pump-probe curves exhibit strong oscillations for high fluence values. For 125 nm Ru film, we did not observe an increase of the signal within the fluence and time delay ranges considered, and thus put the highest measured fluence into the table. We use these data later in \Cref{subsec:a3} when discussing thickness-dependent damage threshold in Ru.\par

\begin{table}[h!]
    \centering
    \begin{tabular}{|c|c|c|c|}
        \hline
        \multicolumn{1}{|p{2cm}|}{\centering thickness, \\ nm} &
        \multicolumn{1}{|p{2cm}|}{\centering incident fluence, \\ mJ/cm$^2$} &
        \multicolumn{1}{|p{2cm}|}{\centering absorbed fluence, \\ mJ/cm$^2$} &
        \multicolumn{1}{|p{2cm}|}{\centering delay, \\ ps} \cr
         \hline\hline
         17 & - & - & - \\
         \hline
         30 & 51.2 $\pm$ 3.8 & 14.1 $\pm$ 1.0 &  25.5 \\
         \hline
         37 & 54.6 $\pm$ 4.0 & 16.7 $\pm$ 1.2 & 36.0 \\
         \hline
         50 & 60.9 $\pm$ 4.5 & 21.7 $\pm$ 1.6 & 21.5 \\
         \hline
         75 & 67.2 $\pm$ 4.9 & 25.9 $\pm$ 1.9 & 21.8 \\
         \hline
         125 & 77.0 $\pm$ 5.7 & 29.9 $\pm$ 2.2 & - \\
        \hline
    \end{tabular}
    \caption{Measured fluence and time delay values related to change of signal trend from slow decay to sharp increase.}
    \label{tab:1}
\end{table}

 The measured data in \cref{fig:prof} show that the peak reflectivity change reached at $\sim 3 \un{ps}$ for each curve increases with the fluence and saturates at a certain fluence value. Similar sets of pump-probe thremoreflectance curves were obtained for various thicknesses of Ru films. The overview of measured data for all film thicknesses is stored in the supplementary materials. \par 
 In \cref{fig:threeR} we provide the peak values of $\Delta R/R_0$ as a function of the pump pulse fluence for three different thicknesses of Ru corresponding to three absorption regimes. The optical penetration depth in Ru for 800 nm light is 16 nm. Therefore, for 17 nm film, multiple reflections of absorbed light at a Ru-Si interface change the absorption profile considerably compared to the Lambert-Beer law. For 37 nm, deviations from the Lambert-Beer profile are visible but not dramatic, whereas 125 nm film optically behaves like a bulk material, see \cref{fig:abs}. We see that with increasing fluence the peak of reflectivity linearly increases and saturates at a particular value for each Ru thickness.\par 

\begin{figure}[h]
    \centering
    \includegraphics[height=0.2\paperheight]{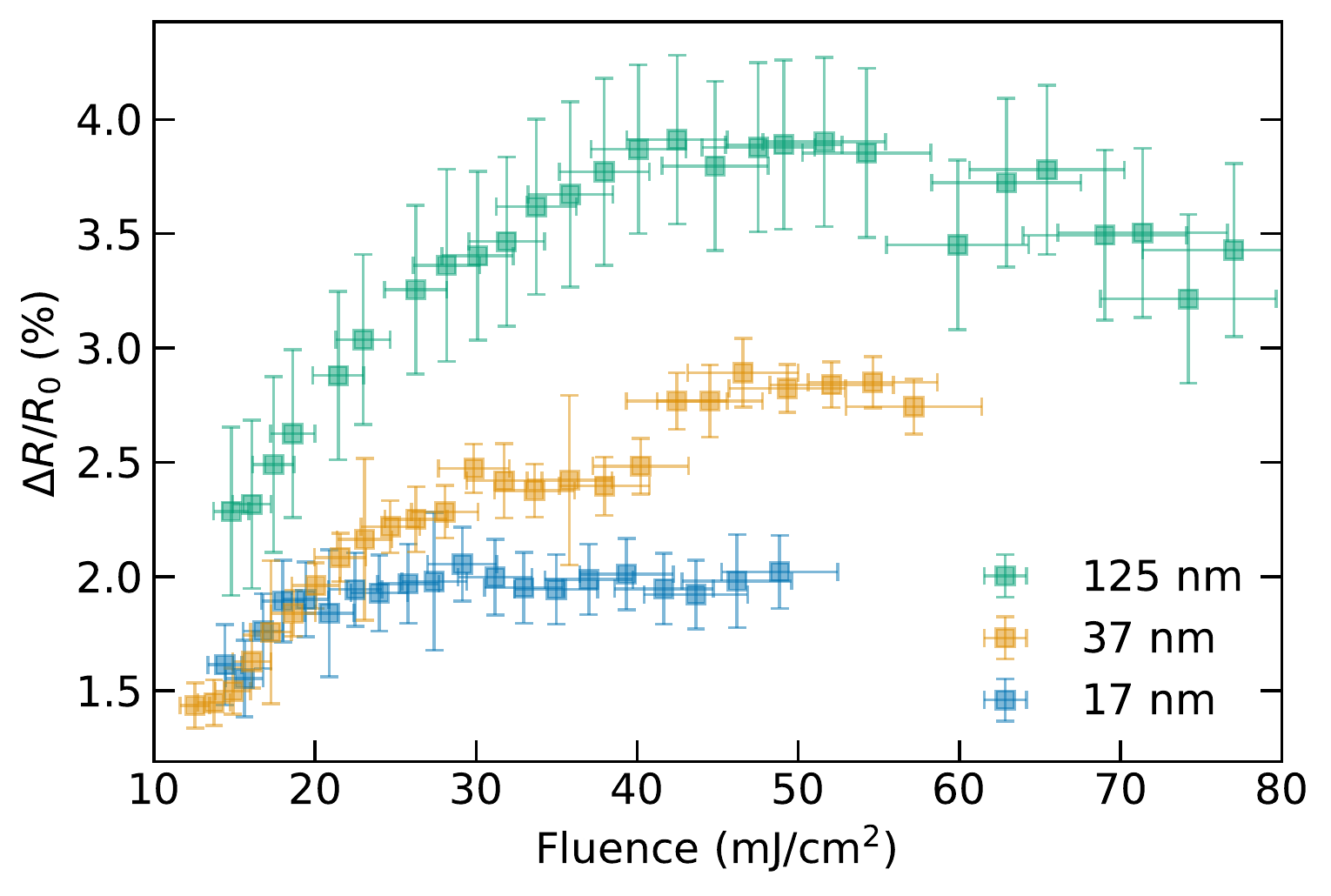}
\caption{Peak values of reflectivity change as a function of pump fluence for 17 nm, 37 nm and 125 nm Ru films.}
\label{fig:threeR}
\end{figure}

\begin{figure}[h]
    \centering
    \includegraphics[height=0.2\paperheight]{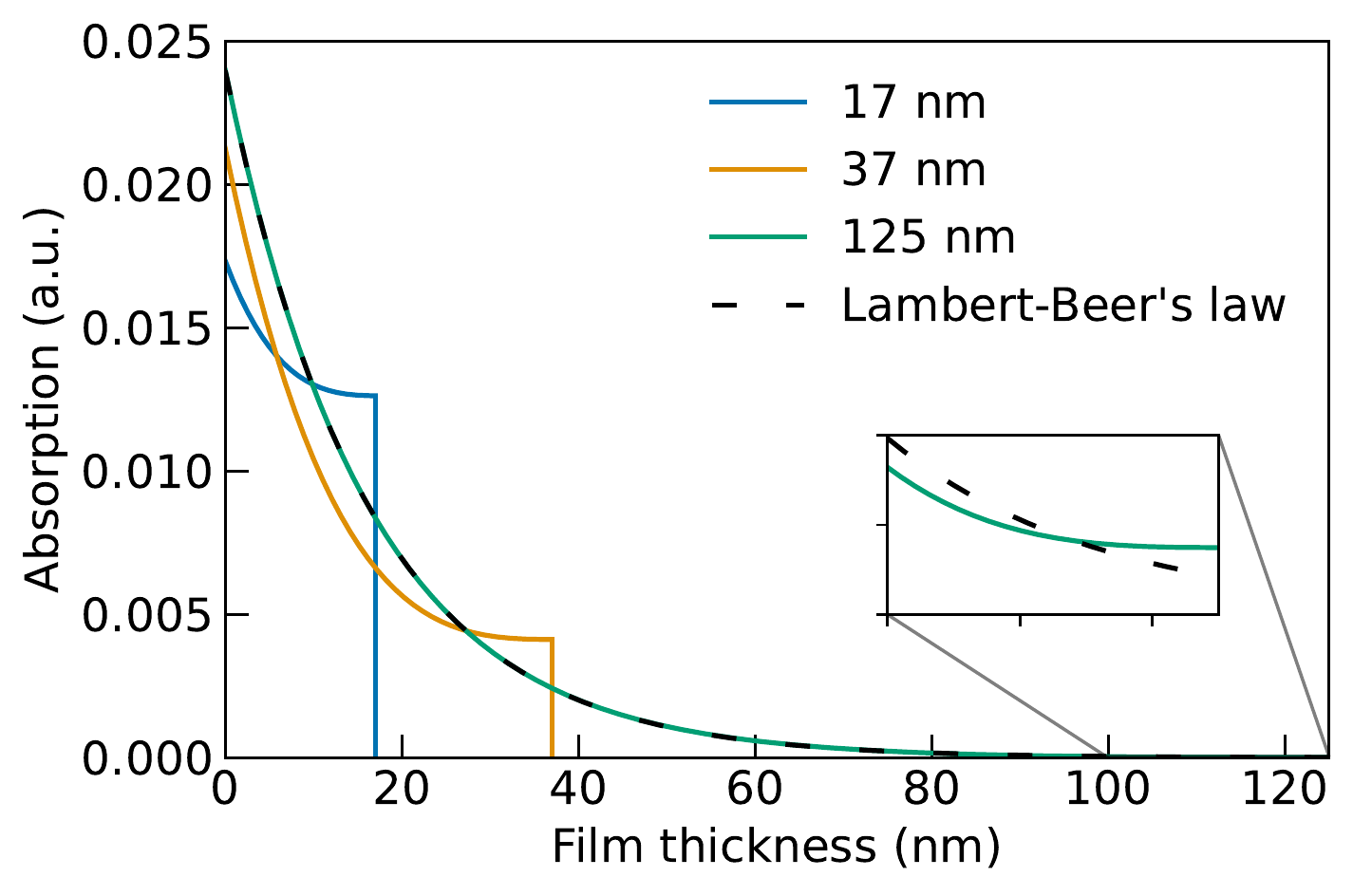}
    \caption{Absorbed energy density profiles for Ru films of 17, 37 and 125 nm thicknesses on Si substrates calculated with the transfer-matrix approach \cite{UDCM-Group2019AbsorptionTMM}. Inset: deviation of transfer-matrix from Lambert-Beer profile at Ru-Si interface.}
    \label{fig:abs}
\end{figure}

In the case of 37 nm Ru, we see a jump in the peak reflectivity values around $45 \un{mJ/cm^2}$. Further analysis does not reveal any sharp changes at this fluence, and such a jump is not present in the data obtained from other samples. Therefore, we consider it to be a measurement artefact. In the case of 125 nm Ru, we observe a drop of reflectivity peak at the highest fluences. We also detected a similar drop at another thickness, 75 nm. We attribute these drops to strong accumulated damage of a film at high fluences. This is confirmed by SEM observations in \Cref{subsec:r2}, where we observed strong damage of sample surfaces at high fluences.\par

\subsection{Analysis of the surface morphology}
\label{subsec:r2}
Since the experiment is carried out with rotation of samples with unsynchronized rotation frequency and laser repetition rate, the pump laser generates not individual spots, but almost continuous lines. Every line corresponds to a particular laser fluence, so we can easily trace possible morphology changes caused by the pump using \textit{ex-situ} SEM analysis.\par
\begin{figure}[h]
     \centering
     \includegraphics[height=0.25\paperheight]{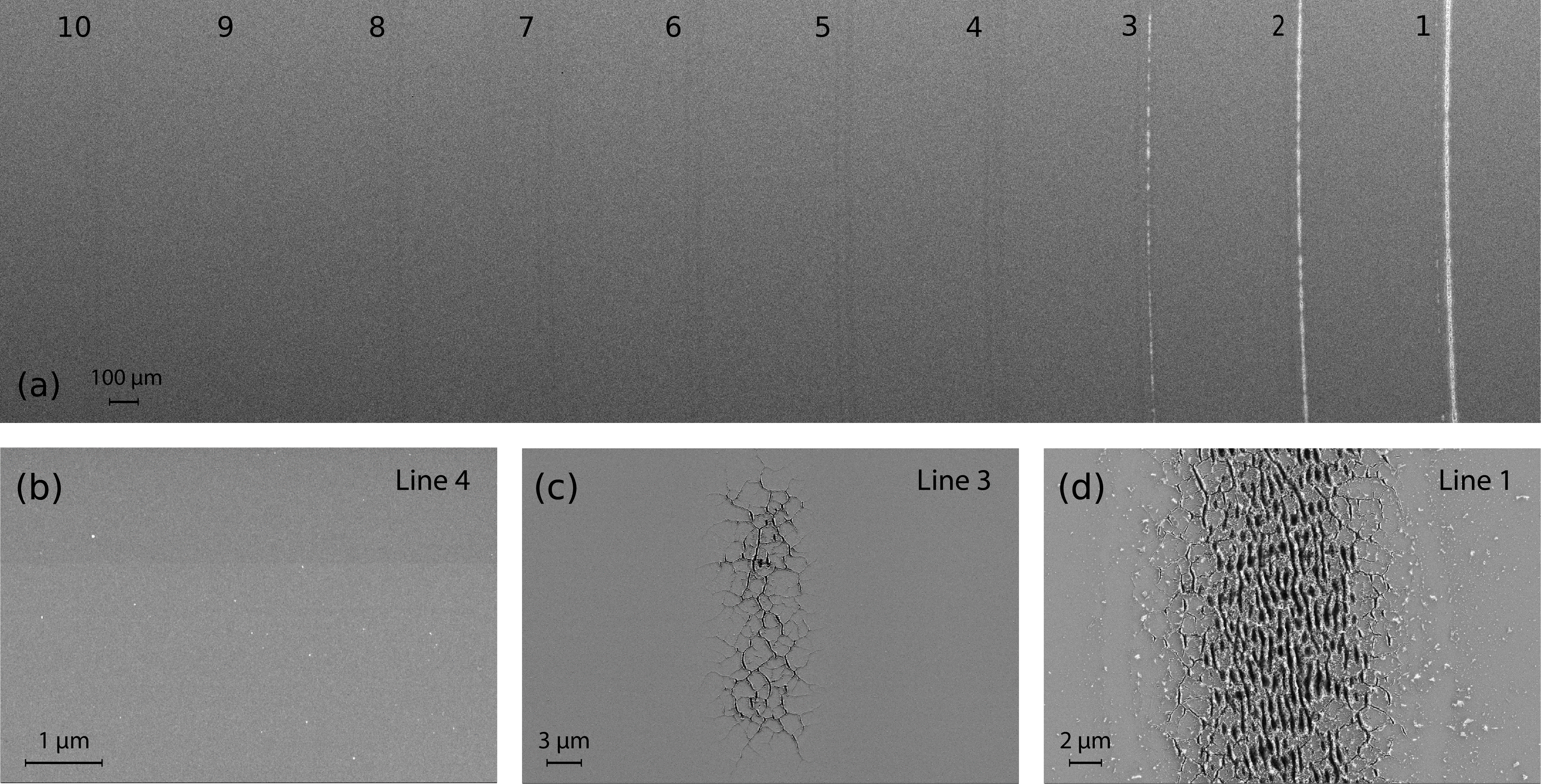}
     \caption{(a) SEM image of 37 nm Ru surface after laser irradiation. Numbered lines correspond to irradiation conditions ranging from the regime of intermediate fluences (lines 10-4, shadows along irradiation path)  to high (lines 3-1, the surface is damaged). (b) Zoomed-in SEM image of line 4, no surface damage is present. (c) The same for line 3. The line consists of islands of cracks along the irradiation path. (d) The same for line 1. Damage morphology is represented by continuous grooves along the irradiation path.}
     \label{fig:sem}
 \end{figure}

In \cref{fig:sem} we show SEM images of a 37 nm thick Ru film surface after laser exposure. \Cref{fig:sem}(a) shows the overview SEM image of exposed lines. The line 1 corresponds to the highest peak fluence of $57 \un{mJ/cm^2}$, the last visible line (line 10) corresponds to the peak fluence of $36 \un{mJ/cm^2}$). Note that before actual damage of the surface, several lines look like shadows. In \cref{fig:sem}(b) we show the magnified view of line 4 (peak fluence is $49 \un{mJ/cm^2}$). At those irradiation conditions, SEM was not able to resolve any morphology changes of the surface. Starting from line 3 we see a clear indication of damage: the surface is covered with cracks (\cref{fig:sem}(c)). 
Further increase of the peak fluence leads to the severe damage of the surface having a shape of periodic grooves. 
 
For the analysis of a shadow on line 4, we performed additional combination of Auger electron spectroscopy (AES), X-ray photoelectron spectroscopy (XPS) and atomic force microscopy (AFM) characterization techniques. AES indicated $\sim 2$ times increase of oxygen concentration on the 'shadow' lines, comparing to reference area outside the exposure zones. A more detailed measurement of the oxide thickness done with angle-resolved XPS indicated only 0.4 nm growth of the oxide layer above the thickness of a native oxide of 0.8 nm. The AFM indicated $\sim 1.5$ times increased RMS roughness in these exposed areas from 0.187 nm to 0.280 nm. It is unlikely that such small structural changes can explain the formation of "shadows" on the SEM image. We can assume possible light-induced carbon growth that could explain such shadows. However, growth of small amounts of C on Ru cannot be reliably quantified by AES and XPS, because carbon and ruthenium emission spectra overlap on Auger electron/X-ray photo-emission spectra. The XPS spectra in the exposed spot showed an increase of the intensity at the energy of the Ru3d$^{3/2}$ peak relative to the Ru3d$^{5/2}$ intensity (compared to outside the exposed area). This is a qualitative indication of an increased carbon content, but the difference compared to the unexposed area is too small for a reliable quantification by peak fitting. Hence, we can claim that the darkness on an irradiated path is a result of heat-induced surface chemistry. We found that the darkness is a result of minor increase of the oxide thickness as well as slight carbonization. However, we could not resolve the exact stoichiometry of the compound since the carbon signal was too weak to quantify.\par

\section{Theoretical analysis}
\label{sec:a}
\subsection{Two-temperature modeling}
\label{subsec:a1}
For the analysis of laser-induced ultrafast heating and melting of Ru thin films, we applied the two-temperature model (TTM) \cite{Anisimov1974}. The TTM equations that govern heat dynamics of electron and lattice subsystems are:
\begin{equation}
\label{eq:2}
    \begin{cases}
        C_e(T_e) \frac{\partial T_e}{\partial t} = \frac{\partial}{\partial z}\left( k_e(T_e, T_l) \frac{\partial T_e}{\partial z}\right) - G(T_e,T_l)(T_e - T_l) + S(t,z),\\
        \left(C_l(T_l)+H_m\delta(T_l-T_m)\right) \frac{\partial T_l}{\partial t} = G(T_e,T_l)(T_e - T_l)\\    
    \end{cases}
\end{equation}
Here $T$, $C$ and $k$ are the temperature, heat capacity and thermal conductivity of Ru electrons (subscript e) and lattice (subscript l), respectively; $G$ is the temperature-dependent electron-phonon coupling factor, and $S$ is the heat source representing a laser pulse. Lattice thermal conductivity is considered to be negligible compared to the electron one. The thermal parameters are provided below in Appendix \ref{app:a1}. To account for solid-liquid phase transition at the melting temperature $T_m$, we used an effective lattice heat capacity containing a delta-function term corresponding to the latent heat of fusion $H_m$, as was initially proposed in \cite{Zhvavyi1997InfluenceHeating} for ns-laser heating, and latter extended in \cite{Bulgakova2005AExplosion} to TTM approach and fs pulses. The melting threshold was considered to be reached when at least one computational cell came to a liquid state. \par
The heat source $S(t,z)$ is a product of a temporal Gaussian pulse and in-depth absorbed energy profile as follows: 
\begin{equation}
\label{eq:3}
    S(t,z) = F \sqrt{\frac{4\ln{2}}{\pi\tau_p^2}} e^{-4\ln{2}(t/\tau_p)^2} A(z) 
\end{equation}
Here $F$ is an incident fluence, $\tau_p$ is a pulse duration, and $A(z)$ is an absorbed energy profile. For bulk materials and films much thicker than the photons absorption length, $A(z)$ can be described by the Lambert-Beer absorption law. However, in the case of multilayer structures and thin films with the thickness comparable to the absorption length of photons, Lambert-Beer law breaks down (see \cref{fig:abs}). To account for the deviations of the absorption profile from the Lambert-Beer's law in the heat source, in our code we implemented the matrix algorithm for the fields calculation \cite{UDCM-Group2019AbsorptionTMM}.
\par

\subsection{Fermi smearing mechanism in Ru}
\label{subsec:a2}
To relate experimentally measured change of reflectivity $\Delta R(t)/R_0$ and TTM-calculated temperatures $T_e(t)$, $T_l(t)$, one needs to know how optical constants depend on temperatures in two-temperature regime. $T_l$-dependence of optical properties is often neglected as far as heating of lattice subsystem does not cause any significant changes in electronic structure. In this work, we assume that the transient optical response of the electron system in Ru can be described in terms of the electron temperature $T_e$, same as in other transition metals \cite{Hohlfeld2000, Norris2003Femtosecondinvited}. However, a seek of an approximate form of $T_e$-dependence of Ru optical constants is a challenging task. Due to the complexity of Ru band structure near the Fermi level, a simple but widely accepted Drude model would not provide reliable results, whereas accurate DFT-based optical constants would require a set of computationally heavy calculations. Nevertheless, some qualitative information about optical response in Ru can be obtained if one recognizes a similarity in band structures of Ru and gold in the vicinity of Fermi level.
\par
\begin{figure}[h]
    \centering
    \begin{subfigure}{0.45\textwidth}
    \includegraphics[width=\linewidth, height=0.25\textheight]{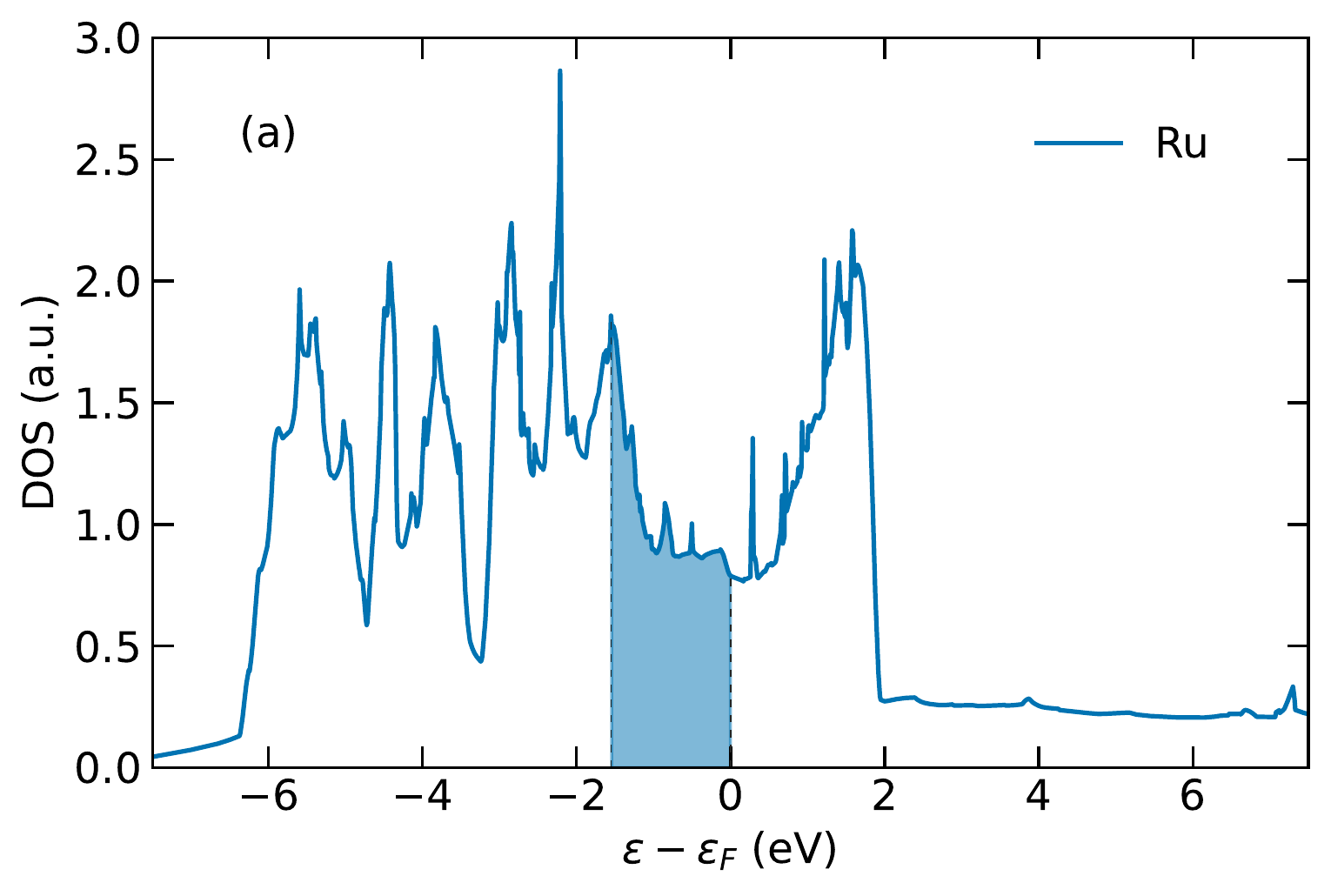}
    \end{subfigure}
    \begin{subfigure}{0.45\textwidth}
    \includegraphics[width=\linewidth, height=0.25\textheight]{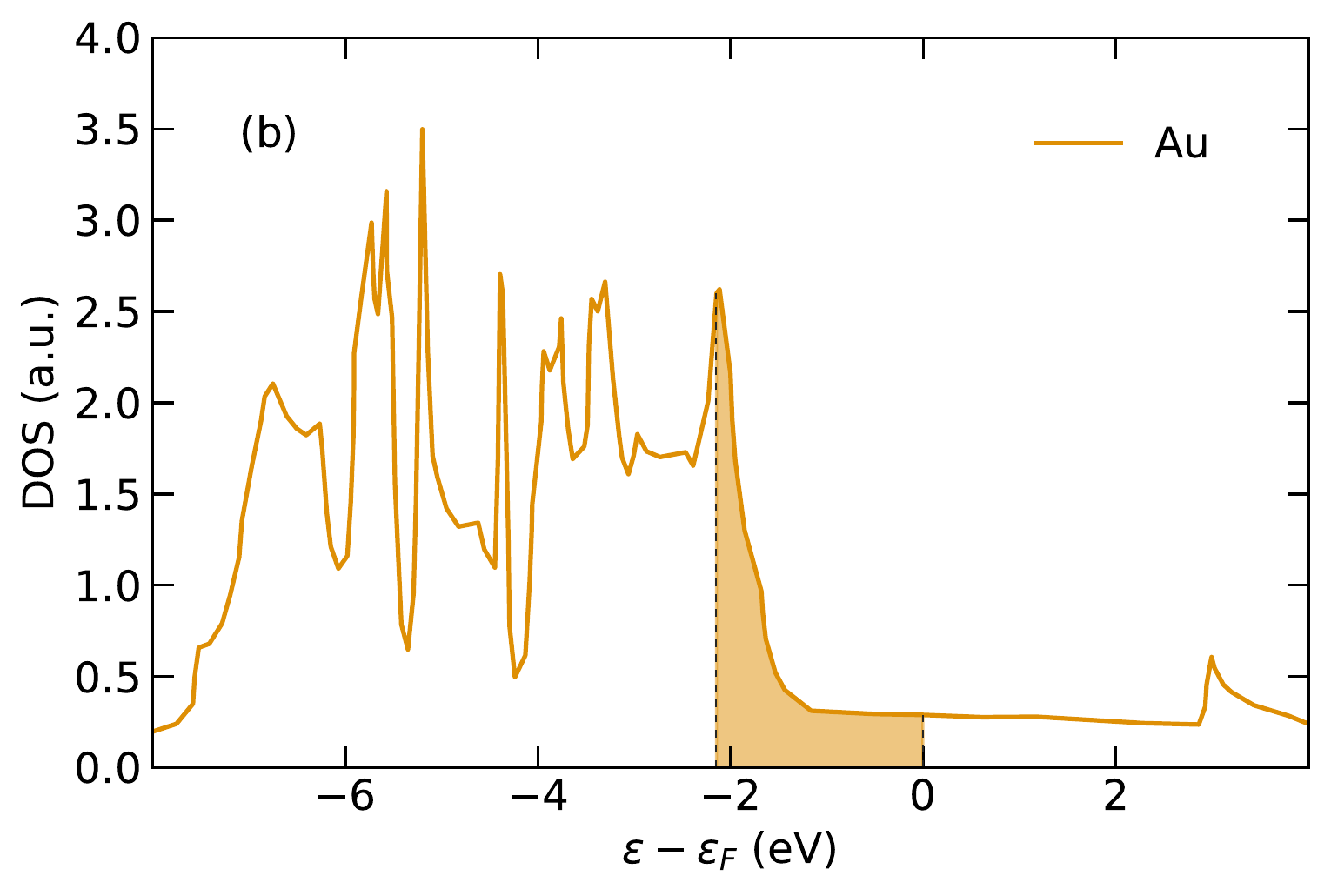}
    \end{subfigure}
\caption{(a) DOS of Ru taken from \cite{Petrov2020} and (b) Au from \cite{Blumenstein2020TransientExperiment}. Blue fill in Ru DOS represents energy states allowing optical transitions for photons with energies $\hbar\omega = 1.55 \un{eV}$, orange fill in Au DOS -- photons with energies equal to d-band transition threshold.}
    \label{fig:dos}
\end{figure}

 
We found out that the electron density of states (DOS) of Ru (\cref{fig:dos} (a)) has a pseudo-gap near the Fermi level ranging from -1.5 to 1 eV, similarly to the gap between the majority of the d-band and the Fermi level in gold (\cref{fig:dos} (b)). As was pointed out in \cite{Bevillon2018NonequilibriumHeating}, for photon energies within this pseudo-gap area, the interband contribution to the optical transitions is weak, and at a qualitative level a temperature dependence of the reflectivity change in Ru may be explained via the Fermi smearing mechanism like in noble metals.
\par
The Fermi smearing mechanism assumes that the reflectivity change is proportional to the change of Fermi distribution with increasing electron temperature, \cref{eq:4} \cite{Lynch1972ThermomodulationCu, Schoenlein1987FemtosecondMetals}.
\begin{equation} \label{eq:4}
    \frac{\Delta R (\hbar\omega,T_e)}{R_0} \sim f(\hbar\omega, T_e) - f(\hbar\omega, T_0),     
\end{equation}
\begin{equation} \label{eq:5}
    f(\hbar\omega, T_e) = \frac{1}{\exp\left(\frac{\hbar\omega+\Delta\varepsilon}{k_B T_e}\right) + 1}.
\end{equation}
Here $\hbar\omega$ is the energy of incindent photons equal to 1.55 eV, $\Delta\varepsilon = \varepsilon - \varepsilon_F$ is the difference between the energy level from which an electron is excited and the Fermi level of Ru, $T_e$ and $T_0$ are the elevated and the initial electron temperatures, respectively. In our case, $\Delta\varepsilon$ is a priory unknown free parameter assumed to be constant for all temperatures, so $\varepsilon$ has a meaning of an averaged energy level participating in optical transition.\par
The Fermi smearing mechanism works in the regime when the photon energy is close to the interband transition energy $\Delta\varepsilon$, which in the case of Ru we attribute to the lower boundary of the pseudo-gap. If we have fixed value $\hbar\omega=1.55 \un{eV}$ and variable parameter $\Delta\varepsilon$, we can try to fit temperature-dependent reflectivity points by \cref{eq:4} with respect to $\Delta\varepsilon$. Then comparing the fitted value with the lower pseudo-gap boundary taken from Ru DOS will allow us to estimate to what extent the Fermi smearing is valid in Ru.\par

\begin{figure}[h!]
    \centering
    \includegraphics[height=0.2\paperheight]{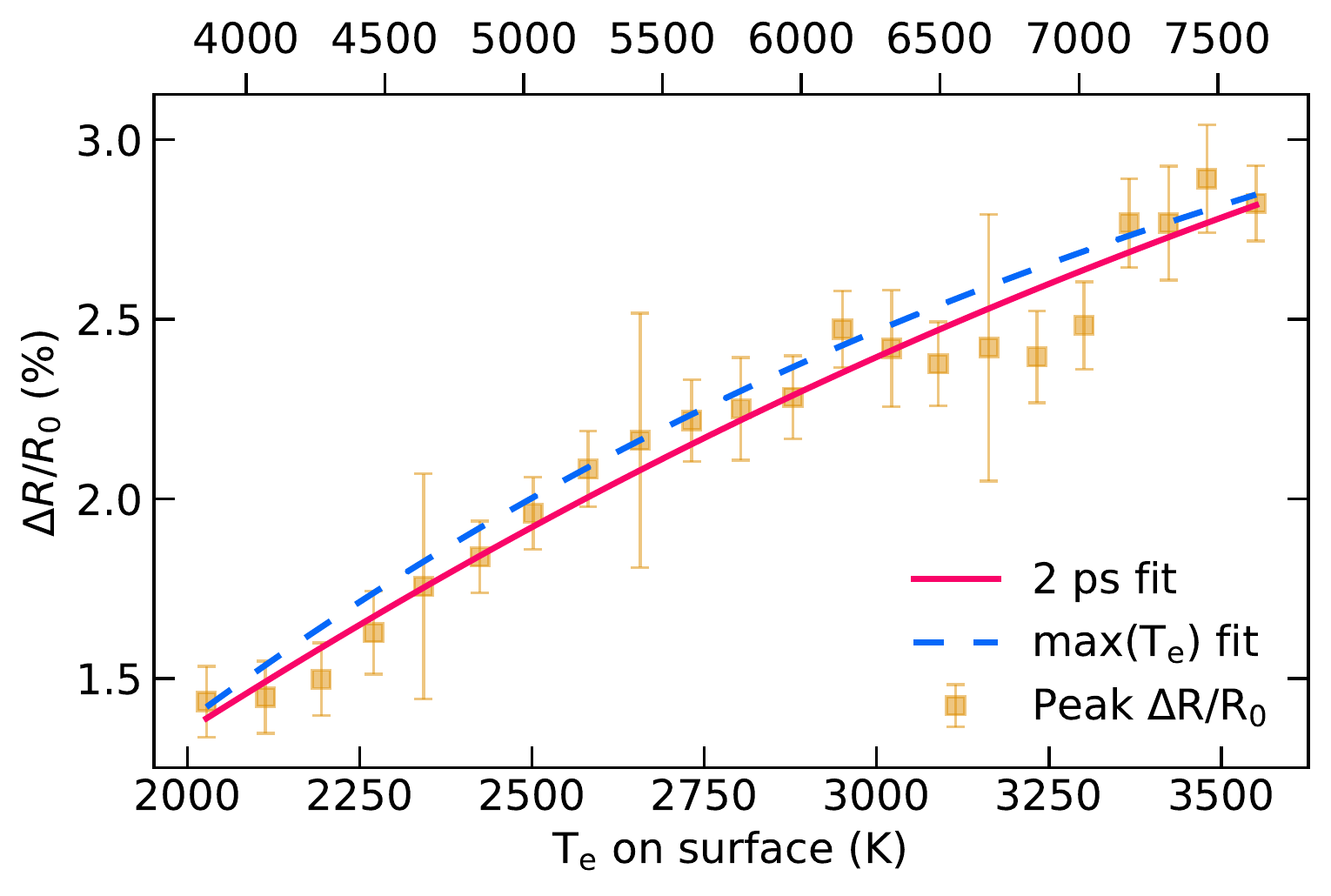}
    \caption{Peak reflectivity change $\Delta R/R_0$ as a function of electron temperature $T_e$ (gray markers). Red line represents fitting by Fermi smearing function \eqref{eq:1} for $T_e$ at 2 ps, and blue dashed line is the same for max $T_e$ value. The ticks on the lower horizontal axis indicate surface temperature values at 2 ps. For the upper axis, ticks correspond to the maximal values of electron temperature.}
    \label{fig:fermi}
\end{figure}

The $T_e$-dependence of the peak reflectivity changes in Ru (\cref{fig:threeR}) can be taken from the TTM model (\cref{eq:2}) with given incident fluence. For every peak reflectivity point we got two temperature values: one is the temperature value at $\sim 2$ ps after the laser pulse as in the real experiment, and the other is the maximal temperature on a surface taken under assumption that the position of the peak reflectivity change corresponds to the maximal heating of the electronic system and not to the process of electron-lattice equilibration. The result of fitting for 37 nm Ru film is shown in \cref{fig:fermi}.
\par

 The fitted averaged energy level $\Delta \varepsilon$ turned out to be $-1.20 \un{eV}$ for $T_e$ at 2 ps. This value is close to the expected -1.5 eV, and thus may indicate the validity of Fermi smearing explanation of the electron temperature dependence of Ru optical properties. For the max $T_e$, the $\Delta \varepsilon$ value is $-0.97 \un{eV}$, which is quite far from the lower boundary of the pseudo-gap. This deviation might confirm our suggestion to attribute the peak reflectivity to the electron-lattice equilibration (reaching of maximal lattice temperature, see \cref{fig:prof}). Nevertheless, the applicability of Fermi smearing mechanism to Ru seems to be limited since the interband contribution to optical transitions is weak but non-vanishing. A more accurate quantitative explanation requires first-principles simulations of Ru in similar manner to \cite{Bevillon2018NonequilibriumHeating} and will be provided in the future works. 

\subsection{Damage threshold in Ru}
\label{subsec:a3}
In this section, we investigated the thickness-dependent damage thresholds. Having the SEM images (similar to \cref{fig:sem}) for all of the samples, we were able to extract thickness-dependent data by attributing the onset of damage to surface cracking. To elaborate on the mechanism of damage we start our analysis with calculations of laser-induced melting threshold. \par
\par
\par
\begin{figure}[h]
    \centering
    \includegraphics[height=0.2\paperheight]{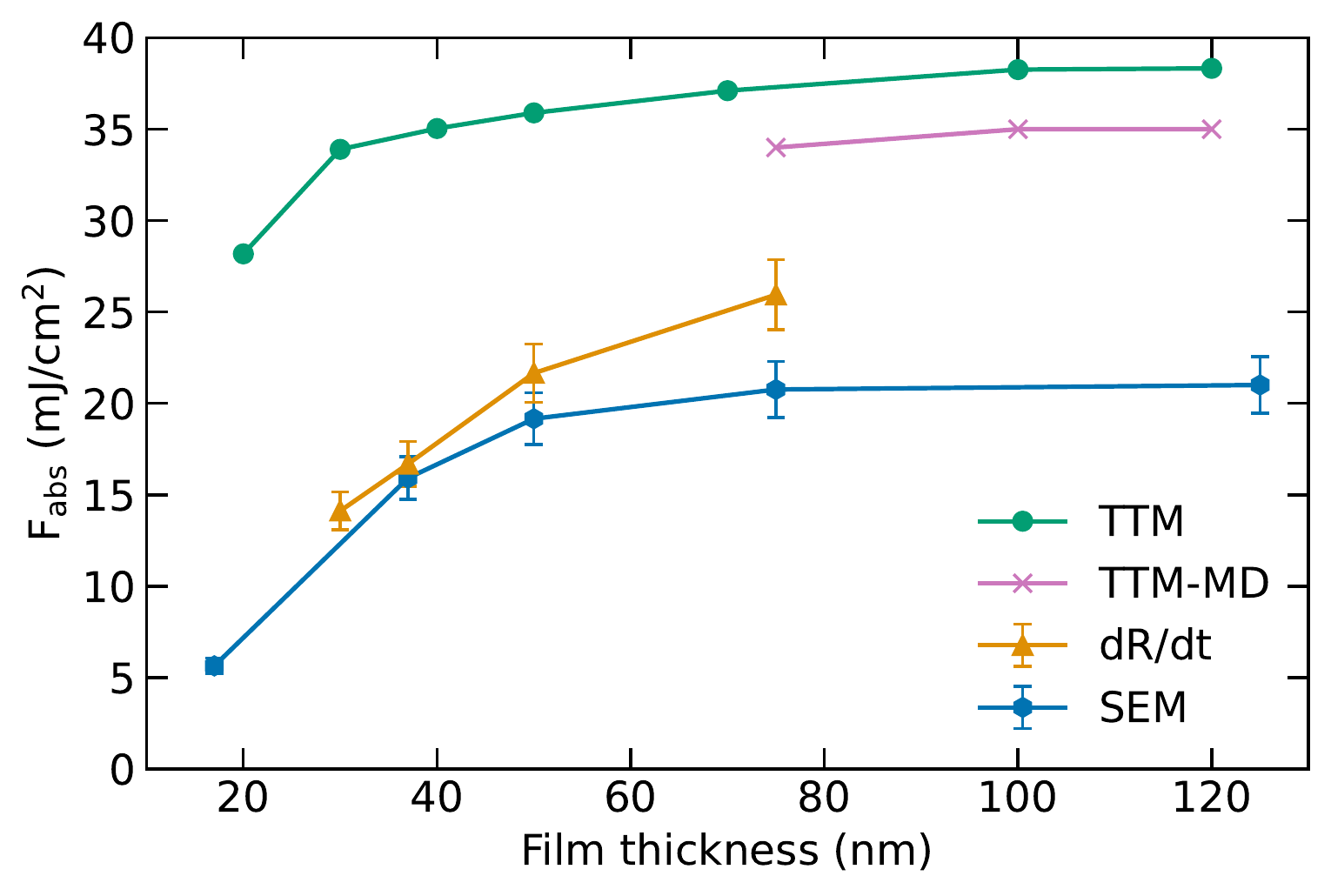}
    \caption{Thickness-dependent threshold fluences of Ru films. 
    Blue hexagons are values corresponding to the first SEM-observed cracks, orange triangles correspond to fluences when pump-probe curves exhibit uncontrolled increase at long timescales (\cref{tab:1}),  green circles are the surface melting threshold provided by TTM simulations, and black cross is the threshold taken from TTM-MD simulation of 100 nm Ru film. 
    }
    \label{fig:th}
\end{figure}

We provided the calculating of melting threshold in Ru irradiated by fs laser using two approaches. The first is a series of TTM simulations (\cref{subsec:a1}) for the different Ru thicknesses. The second is the two-temperature -- molecular dynamics (TTM-MD) simulation for 75, 100 and 120 nm Ru films to stay far from the effects of multiple reflections in absorption profile. In both performed approaches, the melting threshold was determined as fluence at which a surface layer of $\sim 1 \un{nm}$ becomes liquid. In TTM simulations a layer is considered to be liquid when its lattice temperature reaches $T_l=T_m+H_m/C_p$ required for complete melting of the layer, where $H_m$ is enthalpy of melting, and $C_p$ is heat capacity. In TTM-MD we trace a modified centrosymmetric parameter $C_s$ \cite{Murzov2021ElastoplasticWaves} averaged for atoms within the surface layer. It is known that the perfect hcp crystal has $C_s = 3$, whereas melting happens when $C_s$ drops below $2.5$, which is assumed to be a threshold for disordered atom configurations in a liquid phase.  
The results of simulations and their comparison with SEM-thresholds are shown in \cref{fig:th}.

The discrepancy between threshold taken from SEM and fluences when pump-probe singal starts to increase indicate the pump-probe signal stays stable under minor surface changes such as cracks formation. Meanwhile, a factor two difference between experimentally determined threshold and theoretically predicted melting threshold most likely indicates that surface cracking occurs before melting and reduces threshold fluences below single-shot melting. The surface cracking can be caused by generation of heat-induced shear stresses in thin film. The critical shear stress may be estimated from TTM-MD calculation done for absorbed fluence equal to $20\un{mJ/cm^2}$ and is about 6-8 GPa in a surface layer. This amplitude may be attributed to the shear strength of ruthenium dioxide within a thin surface layer, which is comparable with the shear strengths of ceramics damaged by shock loading \cite{Grigoryev2022}. 

Albeit the theoretically predicted and experimentally observed thickness-dependent thresholds seem to attribute to different damaging mechanisms, the follow a common trend. Thresholds increase together with an increasing thickness of a film until some critical value around 75 nm after which they saturate. This critical thickness is associated with change of absorption from thin-film regime to bulk described by Lambert-Beer's law. \par 
Higher threshold values in the TTM than in the TTM-MD can be explained by the absence of the real surface in the TTM. The latent heat of fusion used in the TTM is taken for the bulk material, and thus does not reproduce the properties of the surface. In contrary, molecular dynamics natively takes into account the weaker bonding of surface atoms and thus describes the process of surface melting in a more natural way. It uses the EAM potential constructed specifically for Ru under intense ultrafast laser excitation \cite{Milov2020Two-levelIrradiation}. The melting temperature predicted by this potential is 2787 K, which is 7 \% higher than the experimental value being 2607 K. This leads to a few percent higher threshold values than one would expect from the experiment, but does not have a major impact on the obtained results.

\section{Conclusion}
We presented measurements of transient pump-probe thermoreflectance in Ru thin films as a function of incident fluence and Ru layer thickness in near-threshold regime. We applied rotational scheme to reduce heat accumulation in a target. An analysis of the measured thermoreflectance signal allowed us to extract information about behavior of hot electrons as well as the threshold of irreversible changes in Ru thin films. 
\par
The results of hot electrons analysis indicated similarity of electron system response to laser irradiation for noble metals with fully occupied d-bands and Ru with half-filled d-band. We attributed this result to the presence of a pseudo-gap in Ru DOS around the energy of incident photons. Inside this pseudo-gap, interband transitions are weak, and Ru response may be qualitatively described via Fermi smearing mechanism. A similar effect is expected for other metals with pseudo-gap in d-band (e.g. Cr, W) and may be a scope for a dedicated research. 
\par
We demonstrated occurrence of three well-separated stages of surface changes during ultrafast laser heating of Ru film: darkening of an irradiation line, surface cracking and grooves formation. Our ex-situ surface analysis associated appearance of the darkness with increasing oxidation or carbonization of the surface, however, we were unable to discriminate between the two possibilities. 
\par
We compared the cracks formation threshold to the two-temperature and molecular dynamics simulations of the melting threshold. We found that the crack formation threshold is two times lower than theoretical predictions for single-shot melting threshold. This led us to the conclusion that the basic mechanism of cracking is formation of heat-induced stresses on a thin film surface. Our results may serve as the upper limit of operational conditions for optic devices based on Ru thin films.

\section{Acknowledgements}
FA is grateful to H. van der Velde for fruitful discussions and technical assistance. AK and SS are grateful to C. Berkhout and A. Toonen for technical support.
FA, IMi, JS, IMa and MA acknowledge the Industrial Partnership Program ‘X-tools’, project number 741.018.301, funded by the Netherlands Organization for Scientific Research, ASML, Carl Zeiss SMT, and Malvern Panalytical.\par
For XTANT calculations, computational resources were supplied by the project "e-Infrastruktura CZ" (e-INFRA LM2018140) provided within the program Projects of Large Research, Development and Innovations Infra-structures. NM gratefully acknowledges financial support from the Czech Ministry of Education, Youth and Sports (Grants No. LM2018114, LTT17015 and No. EF16\_013/0001552).  

\begin{appendices}
\renewcommand{\theequation}{\thesection.\arabic{equation}}
\renewcommand{\thefigure}{\thesection.\arabic{figure}}
\section{Two-temperature parametrization of Ru}
\label{app:a1}
\setcounter{equation}{0}
\setcounter{figure}{0}
For electron-phonon coupling factor in Ru, we used parametrization provided by XTANT calculations \cite{Medvedev2020} based on the nonadiabatic tight-binding molecular dynamics approach. In such a scheme, electron-phonon coupling depends on the electron and lattice temperatures $T_e$ and $T_l$ (as well as atomic structure and density). As has been shown in \cite{Medvedev2020}, the dependence on $T_l$ is approximately linear until $T_e$ exceeds some value $\sim 10^4 \un{K}$. In the case of Ru, the dependence is illustrated in \cref{fig:a31} (a) and may be parametrized as follows:
\begin{equation}
    \label{eq:a31}
    G(T_e,T_l) \approx G(T_e) \left(1 + 0.55 \left(\frac{T_l}{T_0} - 1\right)\right) 
\end{equation}
Here $T_0 = 300 \un{K}$ is the lattice temperature at normal conditions (room temperature).\par
Electron heat capacity $C_e(T_e)$ was also extracted from XTANT simulations, see \cref{fig:a31} (b). It demonstrated independence of the lattice temperature, at least below the melting point. The electron thermal conductivity $k_e(T_e,T_l)$ was taken from \textit{ab initio} calculations \cite{Petrov2020}, and that is the most accurate data available to the best of our knowledge.

\begin{figure}[h]
    \centering
     \begin{subfigure}{0.45\textwidth}
    \includegraphics[width=\linewidth, height=0.25\textheight]{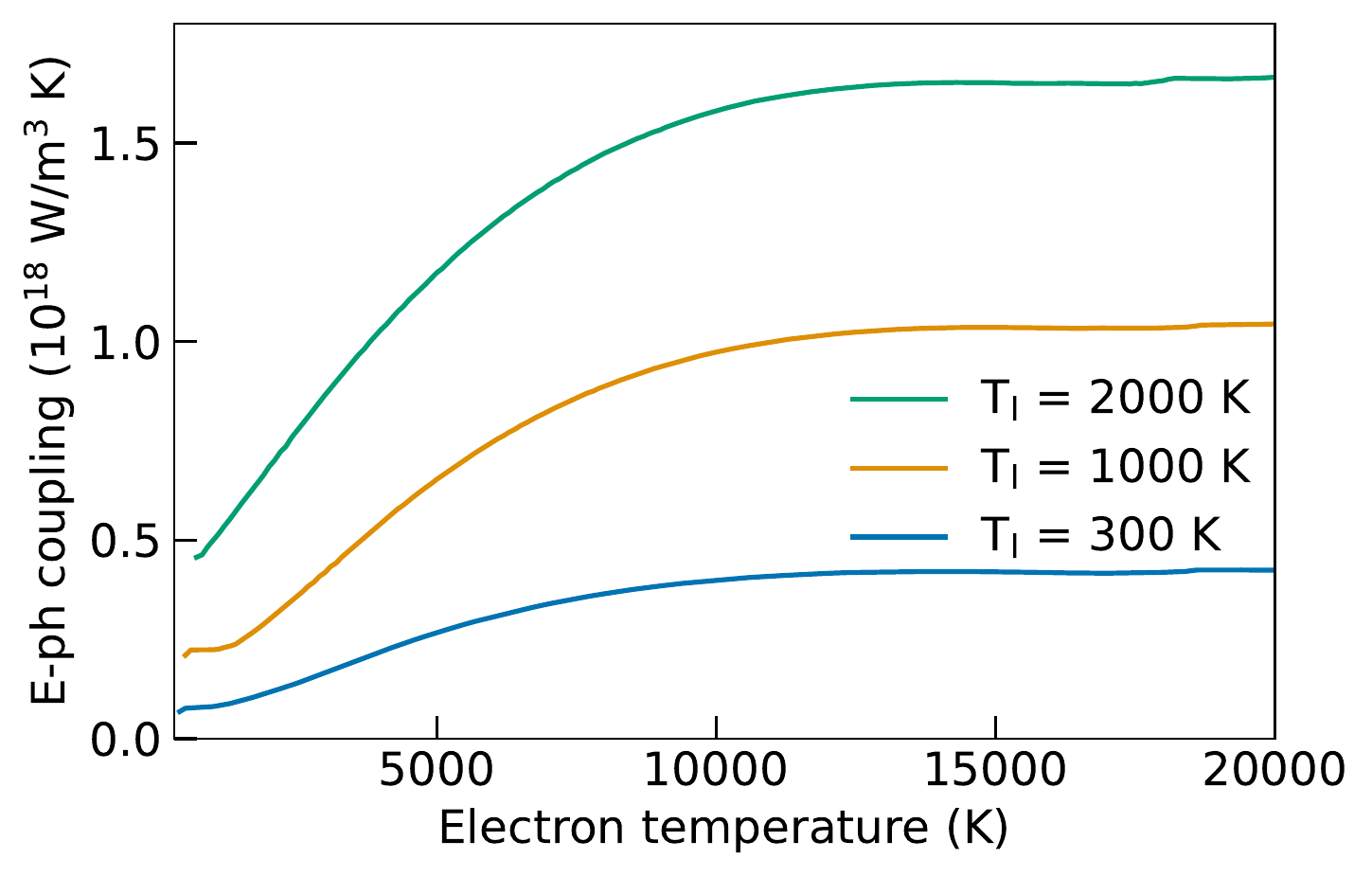}
    \end{subfigure}
     \begin{subfigure}{0.45\textwidth}
    \includegraphics[width=\linewidth, height=0.25\textheight]{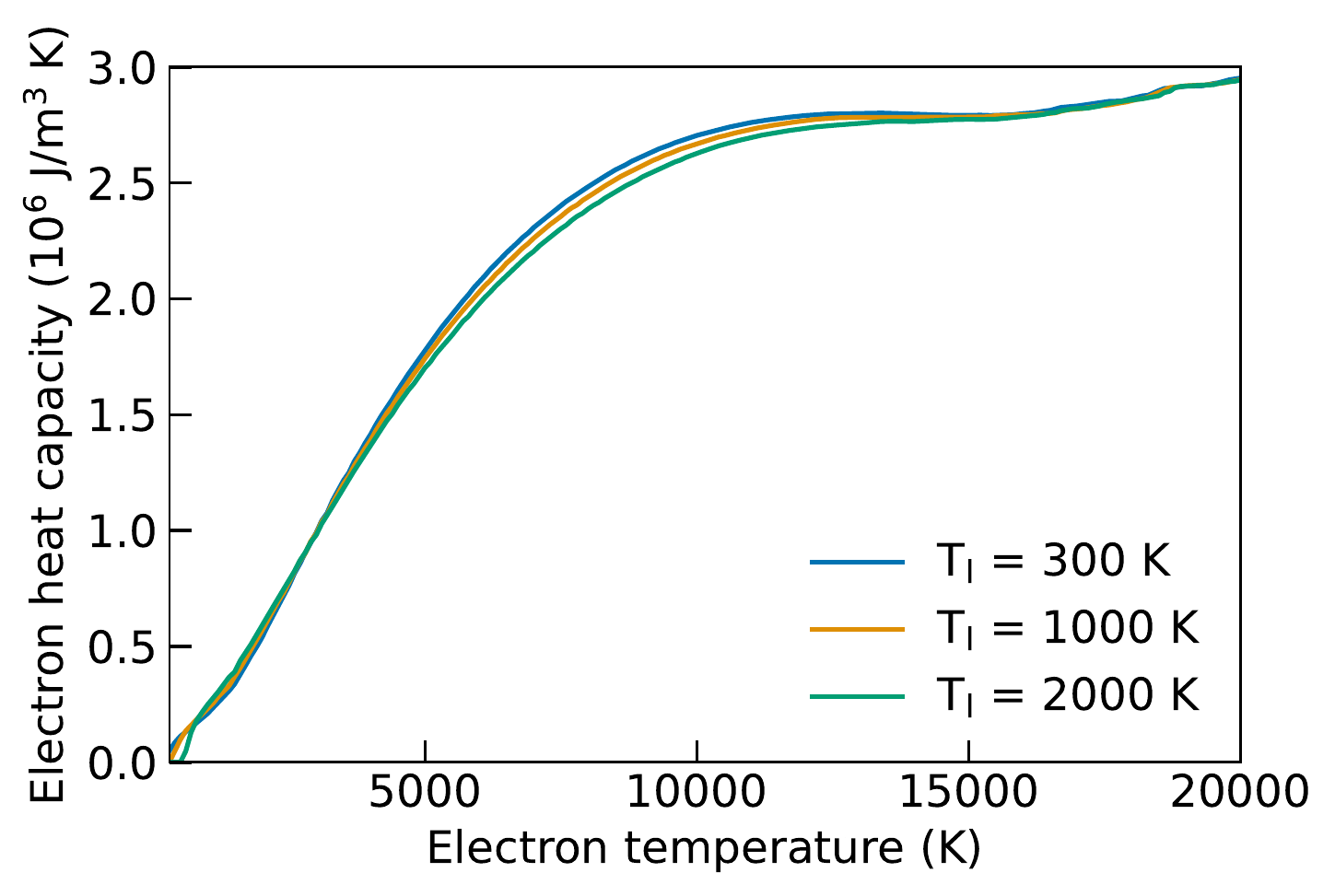}
    \end{subfigure}
\caption{(a) Electron-phonon coupling factor in Ru for different lattice temperatures. (b) Electron heat capacity in Ru for different lattice temperatures.}
    \label{fig:a31}
\end{figure}

\section{Periodic features on pump-probe reflectivtiy curves}
\label{app:a2}
\setcounter{equation}{0}
\setcounter{figure}{0}
Fluence-dependent curves exhibit oscillations especially pronounced at low fluences where signal is less noisy, approximately indicated with dashed lines in \cref{fig:a4}. Such a behavior of pump-probe reflectivity was previously reported in other works, e.g. \cite{Tas1992NoninvasiveInterface, Lin1993StudyUltrasonics}, and is caused by a stress wave generated upon ultrafast heating with a laser pulse. The wave travels back and forth in Ru reflecting at the interfaces with air and Si and partially transmitting into Si (see Fig. 4 in \cite{Milov2020Two-levelIrradiation} as an example of stress wave dynamics in Ru). The amplitude of such a wave is decreasing after each transmission into Si due to the difference in acoustic impedance between Ru and Si, which is confirmed with disappearance of dips after a few travels. Each time the pressure wave comes to a front surface, it changes the temperature and density of Ru, hence transiently changing the measured reflectivity. The period can be expressed as a ratio of the distance that the stress wave travels (two film thicknesses) and a speed of sound in Ru. Therefore, a speed of sound can be estimated from the measured periods. For 37 nm thick film period is $T \approx 11.5 \un{ps}$, and the corresponding estimation of the speed of sound value is $c_s = 2d/T = 6435 \un{m/s}$, what is close to the reported value 5970 m/s \cite{samsonov2012handbook}.

\begin{figure}
    \centering
    \includegraphics[height=0.2\paperheight]{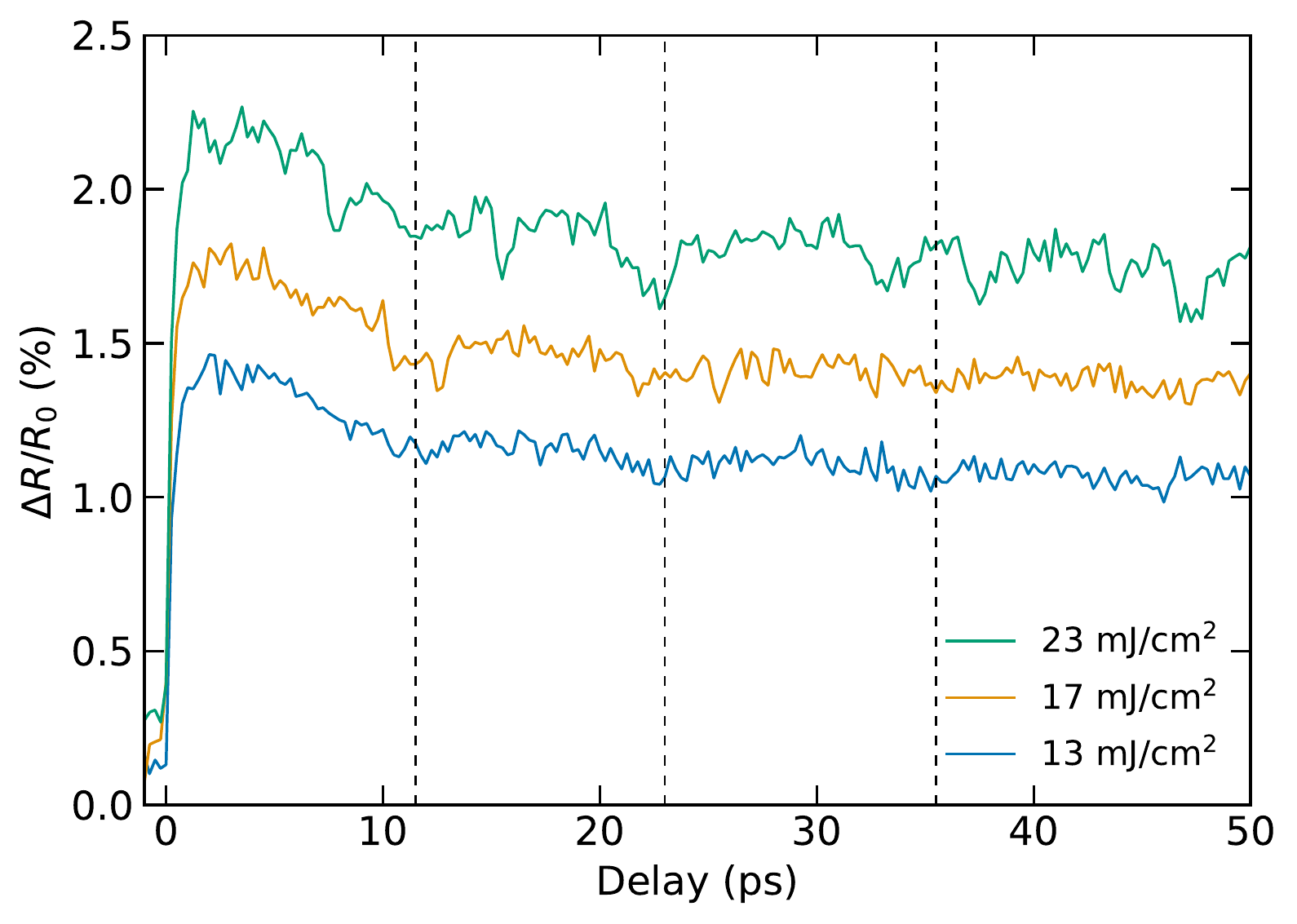}
    \caption{Transient changes in reflectivity of a 37 nm Ru film measured for three incident fluences. The nodes of oscillations indicated with black dashed lines.}
    \label{fig:a4}
\end{figure}
\end{appendices}

\bibliographystyle{unsrt}
\bibliography{refs}
\end{document}